\documentclass[preprint,aps,prb,citeautoscript]{revtex4-1}
\usepackage{graphicx}
\usepackage{siunitx}

\usepackage{color}

\graphicspath{{./figures/}}

\usepackage{mathtools} 

\newcommand{\etal}{\emph{et al.}}

\begin{document}

\title{All-optical switching in granular ferromagnets caused by magnetic circular dichroism}

\author{Matthew O. A. Ellis}
\email{maellis@tcd.ie}
\altaffiliation[Current address: ]{School of Physics and CRANN, Trinity College Dublin, Dublin 2, Ireland}
\affiliation{Department of Physics, University of York, York, YO10 5DD, United Kingdom}

\author{Eric E. Fullerton}
\affiliation{Center for Memory and Recording Research, University of California San Diego, La Jolla, CA 92093-0401, USA.}

\author{Roy W. Chantrell}
\affiliation{Department of Physics, University of York, York, YO10 5DD, United Kingdom}



\begin{abstract}
Magnetic recording using circularly polarized femto-second laser pulses is an emerging technology that would allow write speeds much faster than existing field driven methods. 
However, the mechanism that drives the magnetization switching in ferromagnets is unclear. Recent theories suggest that the interaction of the light with the magnetized media induces an opto-magnetic field within the media, known as the inverse Faraday effect. Here we show that an alternative mechanism, driven by thermal excitation over the anisotropy energy barrier and a difference in the energy absorption depending on polarization, can create a net magnetization over a series of laser pulses in an ensemble of single domain grains. Only a small difference in the absorption is required to reach magnetization levels observed experimentally and the model does not preclude the role of the inverse Faraday effect but removes the necessity that the opto-magnetic field is 10’s of Tesla in strength.

\end{abstract}

\maketitle

\thispagestyle{empty}


The ultimate switching speed of magnetic materials has long been a subject of considerable interest and debate. 
An alternative to switching using magnetic field pulses was demonstrated by Stanciu \etal\cite{Stanciu2007}, who showed that the all-optical control of the magnetic orientation in the amorphous ferrimagnet GdFeCo could be achieved using circularly polarized light pulses. Early explanations by Vahaplar \etal\cite{Vahaplar2009} were based on the inverse Faraday effect\cite{Pitaevskii1961, VanderZiel1965} (IFE) where the polarized light induces a magnetic field within the media. Vahaplar \etal\ showed that the switching by the IFE could be explained by extremely large fields of around 25T. However, it was later discovered that the heating effect of the laser by itself
can cause the magnetization in GdFeCo to switch in an effect known as thermally induced
magnetic switching (TIMS) \cite{Radu2011,Ostler2012}. With this as an underlying mechanism occurring at a critical laser fluence, Khorsand \etal\cite{Khorsand2012} described the helicity dependence arising from magnetic circular dichroism (MCD); in that the media absorbs a different amount of energy from the light depending on the polarity and orientation of the magnetization.

Further experimental\cite{Mangin2014} and theoretical\cite{Evans2014c} work has shown a wider range of materials exhibit all-optical switching in some manner. Intriguingly, helicity dependent switching has been observed in ferromagnetic materials\cite{Lambert2014a} which is unexpected as until now optical control of the magnetization has been confined to ferrimagnetic materials. Specifically it was observed in Co/Pt multi-layers and granular L$1_0$ FePt, both of which exhibit a high anisotropy important for magnetic recording and other magnetic nano-technologies. The underlying physics of TIMS was investigated by Barker \etal\cite{Barker2013}, who showed that the switching is caused by the excitation of two-magnon bound states. The essential requirements for TIMS are; firstly the existence of anti-ferromagnetically coupled sub-lattices and secondly the two sub-lattices must have distinct demagnetization rates, which can be engineered through the dependence on the damping and magnetic moment. With these requirements the ferromagnetic materials in which helicity dependent switching was observed cannot be attributed to TIMS. The opto-magnetic field caused by the inverse Faraday effect remains a possible mechanism but
its precise magnitude and duration are not well understood. 
Since Vahaplar \etal\ found a \SI{25}{T} opto-magnetic field is required to switch GdFeCo one would expect the magnitude required to be much larger to switch FePt due to the greater anisotropy.
Therefore, the underlying mechanism driving helicity dependent switching in ferromagnets is not clear.

In this report we explore and compare two possible mechanisms; first the inverse Faraday effect and second a thermal 'reptation'-like effect. We concentrate specifically on L1$_0$ FePt for comparison with the experiments of Ref.~\onlinecite{Lambert2014a} and is investigated using 
atomistic spin dynamics\cite{Ellis2015_LL, Evans2014} which has proved invaluable 
for investigating both ultrafast laser induced magnetization dynamics\cite{Ostler2012,Barker2013}
and L$1_0$ FePt\cite{Kazantseva2008PRB,Ellis2015_FePt}. 
For the inverse Faraday effect, the magnitude and duration of the opto-magnetic field is varied to construct the regime in which switching is possible.  
Following this we present an alternative, or additional, switching mechanism; a thermal 'reptation'-like model where switching occurs through thermal activation of the grains in analogy to N\'eel's reptation model of hysteresis behavior. In our case during a single laser pulse the grains will switch thermally and magnetic circular dichroism gives rise to a switching rate dependent on the helicity and grain polarity. Therefore, over a sequence of laser pulses a net magnetization averaged over an ensemble of grains will be induced. We utilize a 2 state Master equation model, using the switching probabilities calculated from the atomistic spin dynamics, to predict the evolution of the magnetization as a function of increasing laser pulses. The reptation model gives rise to magnetization changes which increase with the number of laser pulses and which are in agreement with experiments, suggesting that magnetic circular dichroism is the most plausible origin of the helicity dependent all-optical switching in granular FePt.

\section*{Results}

\subsection*{Switching induced by the inverse Faraday effect}

The inverse Faraday effect has been long been suggested as a mechanism for helicity dependent all-optical switching but despite various theoretical treatments\cite{Pitaevskii1961,Hertel2006,Battiato2014} its exact nature has not been fully understood. Therefore, we begin by investigating the magnitude and duration of the magneto-optic field, generated by the inverse Faraday effect, required to cause switching in FePt grains. 
To determine this we employ atomistic spin dynamics; whereby localized atomic magnetic moments are evolved using the Landau-Lifshitz-Gilbert equation\cite{Ellis2015_LL}. The model is specifically parameterized for L$1_0$ FePt using the Hamiltonian derived by Mryasov \etal\cite{Mryasov2005}, for more detail see the methods section.
The heating effect of the femtosecond laser is incorporated through dynamic electron and phonon temperatures, which are evolved using the two-temperature model\cite{Anisimov1974, Chen2006}. The system is equilibrated to room temperature before the laser pulse is applied. The opto-magnetic field is assumed to couple into the spin dynamics in the same manner as an applied field, that is initially zero but triggers with the laser pulse. The field is taken to have the form of a flat Gaussian with a variable central duration. To determine if switching has occurred, the system is simulated for \SI{10}{ps} past the laser pulse and whether the magnetization has passed the $m_z = 0$ plane is monitored. The large anisotropy in L1$_0$ FePt is sufficient that, despite this short timescale, the magnetization will then relax to the easy axis after the laser pulse. This is repeated 20 times to provide the probability of the grain switching from its original orientation into the reversed state.

\begin{figure}
	\centering
  	\includegraphics[width=\textwidth]{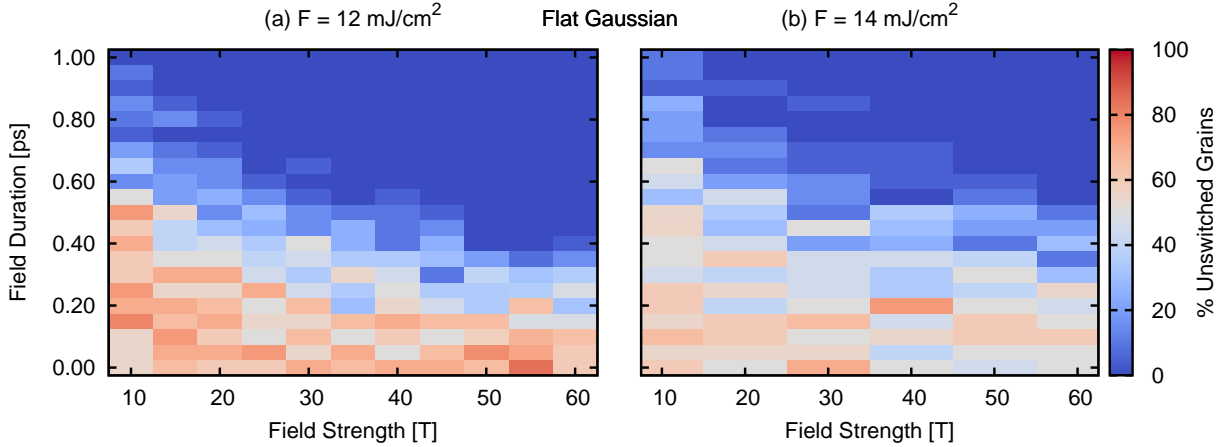}
  	\caption
  	[Percentage of unswitched grains using a flat Gaussian IFE field after
  	a laser pulse of \SI{12}{mJ/cm^2} and \SI{14}{mJ/cm^2}]
  	{\textbf{Computed switching phase space via the inverse Faraday effect.} Percentage of unswitched grains using a laser fluence of (a) \SI{12}{mJ/cm^2} and 
  	(b) \SI{14}{mJ/cm^2} with an inverse Faraday field modeled by 
  	a flat Gaussian for varying strengths and duration. The sides of 
  	the field are Gaussian shaped with a width of $\tau_l =$ \SI{100}{fs}.}
  	\label{fig:Unswitched_gauss_F12_F14}
\end{figure}

 The switching phase space, expressed as the percentage of unswitched grains, is shown in figure \ref{fig:Unswitched_gauss_F12_F14} for laser fluences
of (a) \SI{12}{mJ/cm^2} and (b) \SI{14}{mJ/cm^2}. There is a clear region of deterministic switching for high field strength and long duration for both fluences. However for a \SI{10}{T} field magnitude a duration of approximately \SI{0.6}{ps} is required while for a large \SI{60}{T} field this is only reduced to approximately \SI{0.2}{ps}. This represents the opto-magnetic field remaining in the material in the range of 2 to 6 times longer than the pulse width at magnitudes which are hard to produce externally. With low field magnitude and duration the field is not sufficient to cause consistent switching but there is still a possibility of switching caused by thermal fluctuations yielding a thermally demagnetized film for sufficient laser fluence. Comparing (a) and (b) we can see that the increased fluence has not improved the switching window but rather the enhanced thermal effects causes a more randomized final state. This can be seen as in the low magnitude-short duration regime the probability of remaining unswitched is reaching 50\%.

\subsection*{Thermal Switching Of Grains}
Whilst the opto-magnetic fields required to cause switching through the inverse Faraday effect are large it is clear that the thermal effects can dominate the switching. The effect of the laser heating is to quickly demagnetize the grain within a few hundred femtoseconds but then the re-magnetization process is on the time-scale of picoseconds\cite{Kazantseva2008EPL}. This implies that within a few picoseconds of the laser pulse the magnetization may thermally hop over the anisotropy energy barrier. This energy barrier is reduced due at elevated temperature due to magnetization fluctuations and is naturally included in the atomistic model~\cite{Asselin2010}.

\begin{figure}[b]
  \centering
  \includegraphics[width=0.6\textwidth]{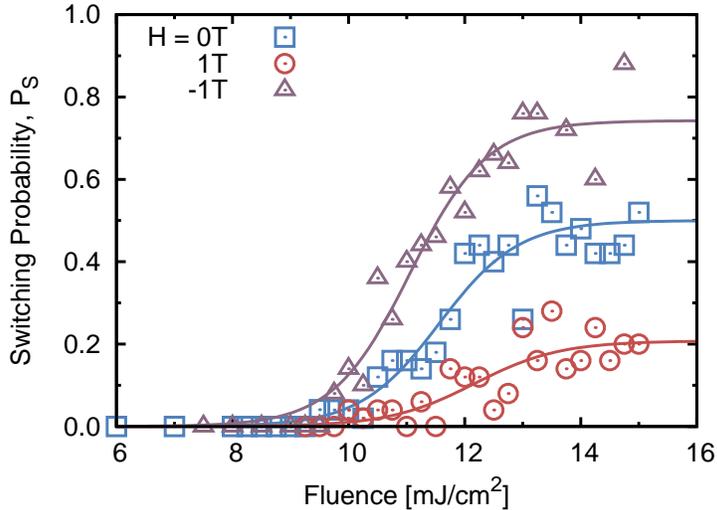}
  \caption
  [Probability of grains thermally switching in zero field and in a constant field 
  of $\pm$ 1 T]
  {\textbf{Computed probability of a single grain reversing over a range of laser fluences.} 
  The probability of a single grain switching without the inverse Faraday effect. 
  Even without a field the grain thermally switches when the fluence is high 
  enough to demagnetize the grain.  Above F = \SI{16}{mJ/cm^{2}} the system is 
  above the Curie temperature and will cool over a longer timescale. The switching 
  is also calculated for a constant field of \SI{1}{T} either aligned or anti-aligned with 
  the initial magnetization direction. The lines show a fit to the data.}
\label{fig:Switching_H0}
\end{figure}

To understand the role that thermal effects play, we now investigate the switching probability, that is the probability of a grain to reverse its polarity during a single laser pulse as a function of the laser fluence without any opto-magnetic field. Figure
\ref{fig:Switching_H0} shows the switching probability averaged over 50 separate pulses in zero field and with a constant $\pm$ \SI{1}{T} external field. As expected, at low fluences, below \SI{8}{mJ/cm^{2}}, the ratio $KV/k_BT$ is large and essentially the switching probability is zero. Above
\SI{8}{mJ/cm^{2}} there is a possibility for the grains to switch direction and the probability increases strongly with fluence until about
\SI{14}{mJ/cm^{2}} where the peak electron temperature is high enough to fully demagnetize the grain. At this point the grains would be evenly distributed between up
and down and so the probability of switching would be 50\%. Beyond \SI{16}{mJ/cm^{2}} both the electron and phonon temperatures remain above the Curie temperature on a much longer timescale governed by the cooling rate of the sample. The effect of a constant applied field is to provides a bias to the energy barrier and thus decreases or increases the switching probability depending on whether it is oriented parallel or anti-parallel to the starting magnetization.

\subsection*{Reptation model of helicity dependent switching}

Within this picture we proceed to derive a simple formalism for the magnetization induced over a sequence of laser pulses. The mechanism of magnetization acquisition is as follows. The laser heats all grains, but those grains oriented in the higher absorption direction will achieve a higher temperature with consequently a relatively large switching probability according to Fig.\ref{fig:Switching_H0}. Any such grains switched will reside in the lower absorption direction, with a reduced probability of switching back. In this picture we expect a greater probability of the grain ending in the lower absorption direction. We note that since the difference in absorption is small the magnetization change in a single pulse will be small. 
However, in the 
experiments performed by Lambert \etal\ the samples were subjected to a sequence of laser pulses at a repetition rate of 1kHz and consequently each grain will be excited by the laser around 1000 to 10000 times. This is expected to lead to a continuous acquisition of magnetization following each laser pulse; an optically-induced reptation, analogous to the magnetization changes accumulating after a number of field pulses in N\'{e}el's classical reptation phenomenon. In comparison to Khorsand \etal{}'s model of helicity dependent switching in GdFeCo; TIMS provides an underlying mechanism and occurs with a single pulse while our model here is driven by thermal switching and a final magnetization state is built over many pulses.

To estimate the induced magnetization caused by thermal switching with magnetic circular dichroism a two-state Master equation is used. Since we are interested in granular L$1_0$ FePt  each grain is a single domain and the due to the large anisotropy the magnetization is fixed out of plane. Therefore we can consider the probability of each grain in the ensemble occupying either out of plane orientation. Due to the  magnetic circular dichroism one of these states will absorb more energy and the other less so we consider the probability of the grain occupying the high or low absorbing orientation; $n_+$ and $n_-$ respectively. For simplicity we neglect inter-granular interactions, which firstly will not affect the underlying physics involved and secondly will be of limited importance since the switching takes place at elevated temperatures.
The induced net magnetization, normalized to the equilibrium value, orientated in the direction of magnetization of the low absorption state is given by 
$ m = n_- - n_+ $. The time evolution of these is given by

\begin{align}
	\frac{\partial n_+}{\partial t} = - n_+ \tau_+^{-1} + n_-\tau_-^{-1} \label{eqn:dnplus} \\
	\frac{\partial n_-}{\partial t} = - n_- \tau_-^{-1} + n_+ \tau_+^{-1} \label{eqn:dnminus}
\end{align}

where $\tau_\pm^{-1}$ are the transition rates of the grain
switching from the high ($+$) or low ($-$) absorption states to the other state. 
Earlier the probability of a grain switching was calculated for a 
single laser pulse; we now consider a repeating laser so that the reptation effect can occur. In this case the time domain is constructed as $t = N / f_l$ with N the number of the  pulse  and $f_l$ the repeat frequency of the laser pulse. Now the transition rates as a function of the laser fluence, F, are described by

\begin{equation}
	\tau^{-1}_\pm(F) = f_l P \left( F \delta_\pm \right),
   \label{eq:ratenew}
\end{equation}
where $P \left( F \delta_\pm \right)$ is the probability of the grain switching by a single laser pulse and $\delta_\pm = 1 \pm \Delta / 2$ is a factor describing the difference in the absorbed fluence due to the magnetic circular dichroism. $\Delta$ is the MCD ratio, which is of the order of a few percent. For a detailed parameterization of this model the switching probability is taken to be that found using the atomistic spin model in Fig.~\ref{fig:Switching_H0}. A function is fitted to the data for each of the different field strengths and a linear interpolation of the fitting parameters is used; more details are provided in the methods section.

In equilibrium the states will satisfy detailed balance; using this condition and the conservation of the total probability, $n_+ + n_- = 1$, we find
\begin{align}
	m^\infty &= \frac{\tau_-^{-1} - \tau_+^{-1}}{ \tau_-^{-1} + \tau_+^{-1}} \\
    		 &= \frac{ P(F\delta_-) - P(F\delta_+) }{ P(F\delta_-) + P(F\delta_+) }
\end{align}
This shows the steady state average magnetization achieved over a series of laser pulses. In essence this relies on the relative difference in the transition rates which can be non-zero even at very low laser fluence.

Solving equations \eqref{eqn:dnplus} and \eqref{eqn:dnminus} the time dependence of the induced magnetization is given by 
\begin{equation}
	m(t) = m^\infty -  (m^\infty  - m(0) )\exp^{-\frac{t}{\tau}}, 
    \label{e:moft}
	\end{equation}
with the magnetization relaxation rate given by

\begin{equation}
\tau^{-1} = \tau_+^{-1} + \tau_-^{-1}.
\end{equation}

\begin{figure}
	\includegraphics[width=0.46\textwidth]{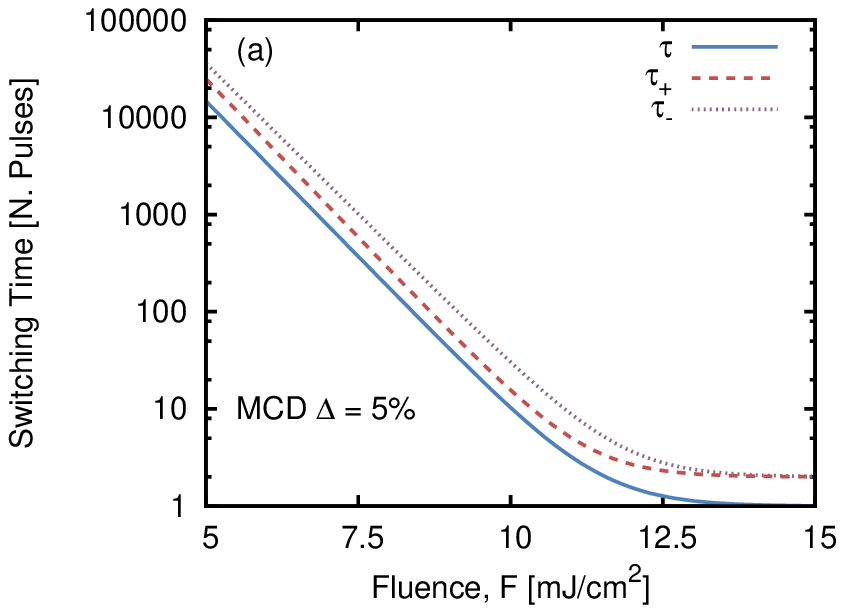}
    \includegraphics[width=0.53\textwidth]{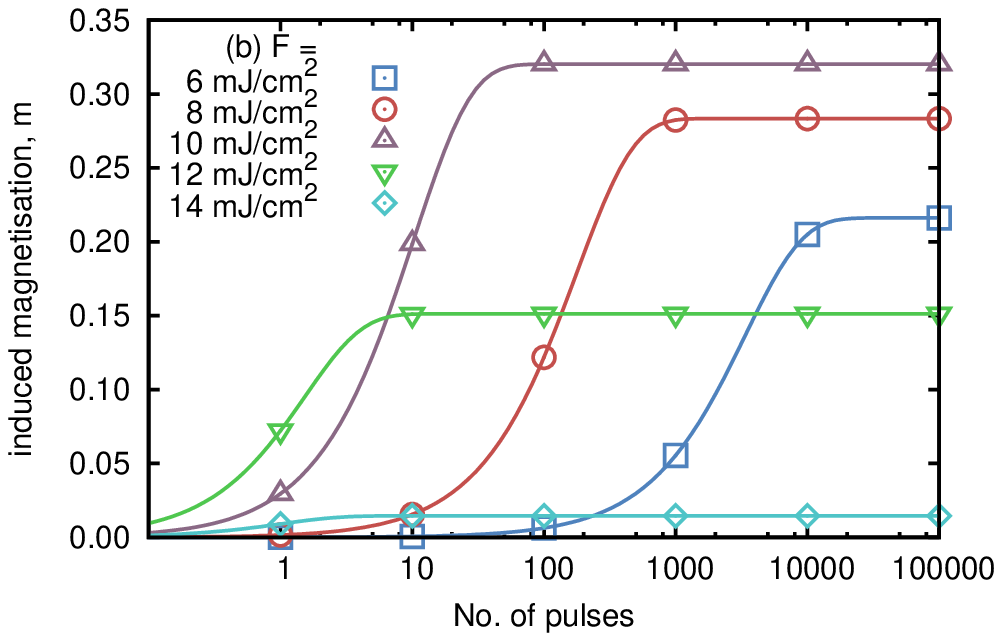}
	\caption{
    \textbf{Switching times and acquisition of magnetization over a series of pulses.}
    (a) The switching times; $\tau$, $\tau_+$ and $\tau_-$ as a function of fluence with a MCD of 5\% expressed as number of laser pulses.
    (b) The induced magnetization with the number of laser pulses for different fluences. At low fluence the switching time is large and so many laser pulses are required to reach the final induced magnetization. At higher fluences the induced magnetization saturates within a few pulses but since thermal effects are higher the induced magnetization is lower.}
    \label{fig:MCD_Mvtime}
\end{figure}

$\tau$ is shown in figure \ref{fig:MCD_Mvtime}.(a), alongside
$\tau_+$ and $\tau_-$ over a range of fluences. $\tau$ shows no
variation with different MCD ratio while $\tau_+$ and $\tau_-$ show a widening separation for larger MCD, here only a relatively large MCD ratio of 5\% is shown. The switching times drop significantly with fluence since the probability of switching shown in figure \ref{fig:Switching_H0} is initially very small but follows a sharp transition to 50\%. 
The time evolution of the induced magnetization is shown in figure \ref{fig:MCD_Mvtime}.(b) for an MCD of $\Delta = 5\%$ calculated from equation \eqref{e:moft}. The net relaxation time depends strongly on the fluence and at low fluence the induced magnetization takes a large number of laser pulses to reach the equilibrium value. At high fluence the equilibrium state is reached in a relatively few laser pulses but the magnitude of the induced magnetization is much smaller since the temperature reached is significantly higher.

Equation \eqref{e:moft} can be solved numerically for any initial magnetization state. The experiments of Lambert \etal\cite{Lambert2014a} were carried out on a demagnetized sample, so we consider an initial state with equal numbers of spins up and down. For this special case with, $m(0)=0$, it is easy to show that the gradient for a small number of laser pulses is
\begin{equation}
\left ( \frac{d m}{d N} \right )_{N\rightarrow 0} = \Delta P,
\label{eq:analytic}
\end{equation}
where $\Delta P = P(F\delta_+)-P(F\delta_-)$. This shows that over a few laser pulses the acquisition of magnetization is essentially linear building up after each step. Equation \eqref{eq:analytic} gives a simple approach to determine the experimental values of $\Delta P$ which, along with the single shot switching probability, either from experiments or atomistic calculation, can be directly related to the MCD value; potentially an important check on the validity of the model.

\begin{figure}
	\centering
    \includegraphics[width=0.45\columnwidth]{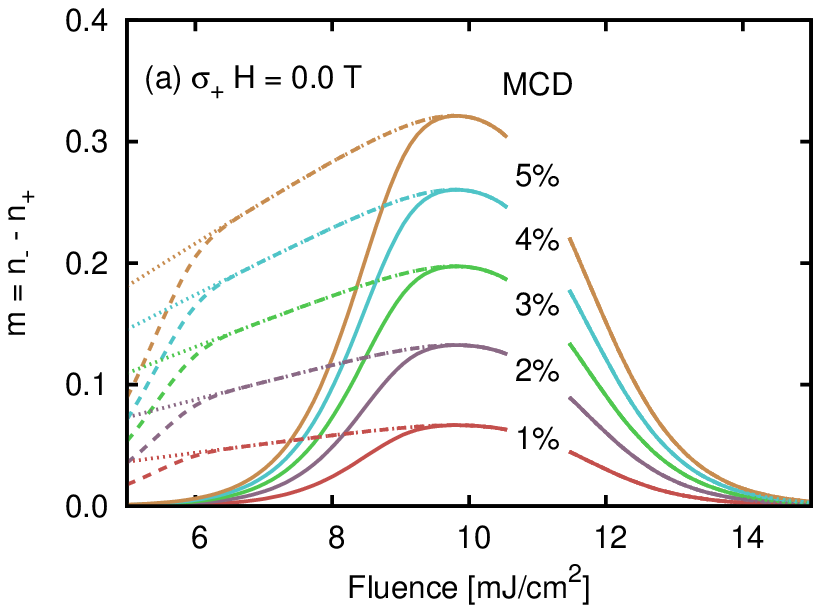}
    \includegraphics[width=0.45\columnwidth]{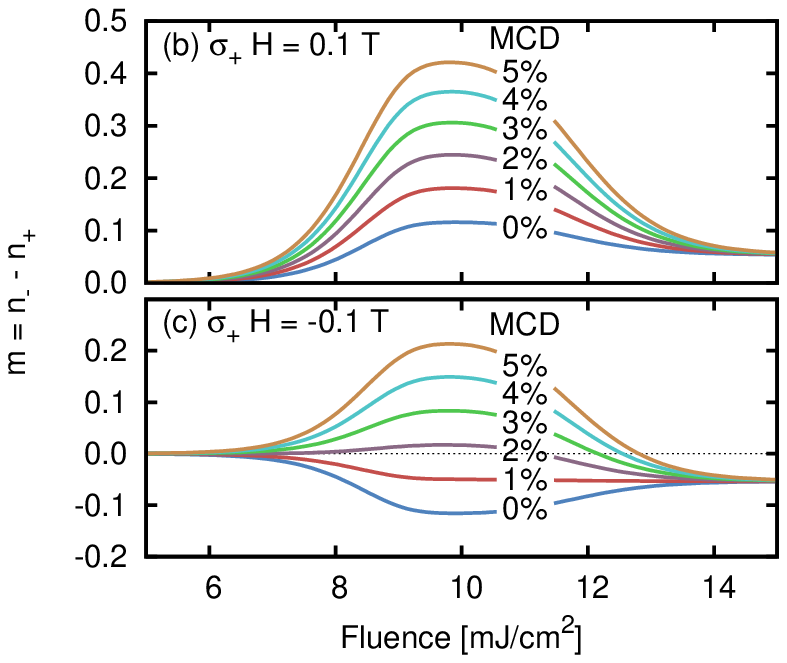}
    \caption{
    \textbf{Induced net magnetization after a series of laser pulses.}
    The induced magnetization due to the MCD effect over a range of fluence and MCD percentage for (a) zero, (b) +\SI{0.1}{T} and (c) -\SI{0.1}{T} constant applied field strengths. The solid lines show the induced magnetization after 100 pulses while in (a) the dashed lines are after 1000 pulses and the dotted lines are $m^\infty$. A peak induced magnetization appears at approximately 9.5 mJ/cm$^2$ can for and MCD ratio of 3\% gives rise to an induced magnetization comparable with experimental measurements. }
    \label{fig:induced_mag}
\end{figure}

The magnetization induced by the MCD is shown in figure \ref{fig:induced_mag} as a function of laser fluence for a range of MCD ratios. The solid lines show the induced magnetization after 100 laser pulses and the dashed lines after 1000 while the dotted lines show the equilibrium magnetization, $m^\infty$. Below a critical value of fluence, no acquisition of the magnetization occurs, essentially because the laser heating is not sufficient to induce thermally activated transitions over the energy barriers. A weak dependence of the critical fluence with MCD ratio may be expected since the high absorption orientation may absorb sufficient energy to become thermally active but since the MCD ratio is small the central fluence needs to be large. Above the critical fluence the behavior is non-monotonic, exhibiting a rapid increase resulting from increased thermal activation. At a fluence of around \SI{9.5}{mJ/cm^2}, independent of the MCD value a peak is reached, after which the magnetization decreases due to the decrease in $m^\infty$ resulting from increasing thermal instability at elevated temperatures.

Whilst this simple model is unlikely to give quantitative results it does show that even a small MCD will give rise to a measurable induced magnetization over a series of laser pulses. The maximum induced magnetization is close to where the steepest gradient of the switching probability but due to the thermal randomization at high fluences it is not centered on the steepest part. The induced magnetization in figure \ref{fig:induced_mag} is defined as in the orientation of the low absorption state and since the values are positive this implies that the system will move towards the low absorption orientation. Khorsand \etal\ measure a MCD of 1.5\% for GdFeCo and also demonstrate that it can be increased by tailoring the structure of the multilayer sample increasing up to 3\%. This implies tailoring the sample would be a possible route in increasing the magnitude of the induced magnetization.
 
\begin{figure}[htb!]
	\centering
  	\includegraphics[width=.35\columnwidth]{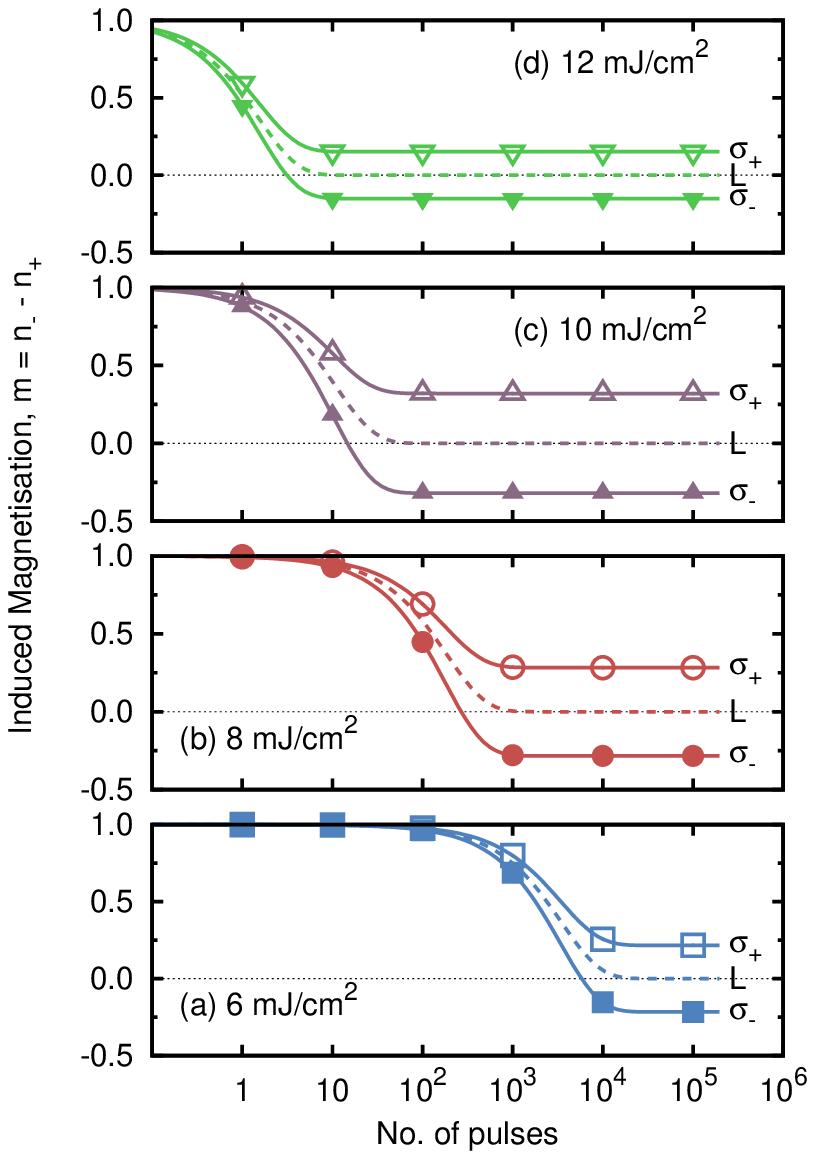}
  	\includegraphics[width=.35\columnwidth]{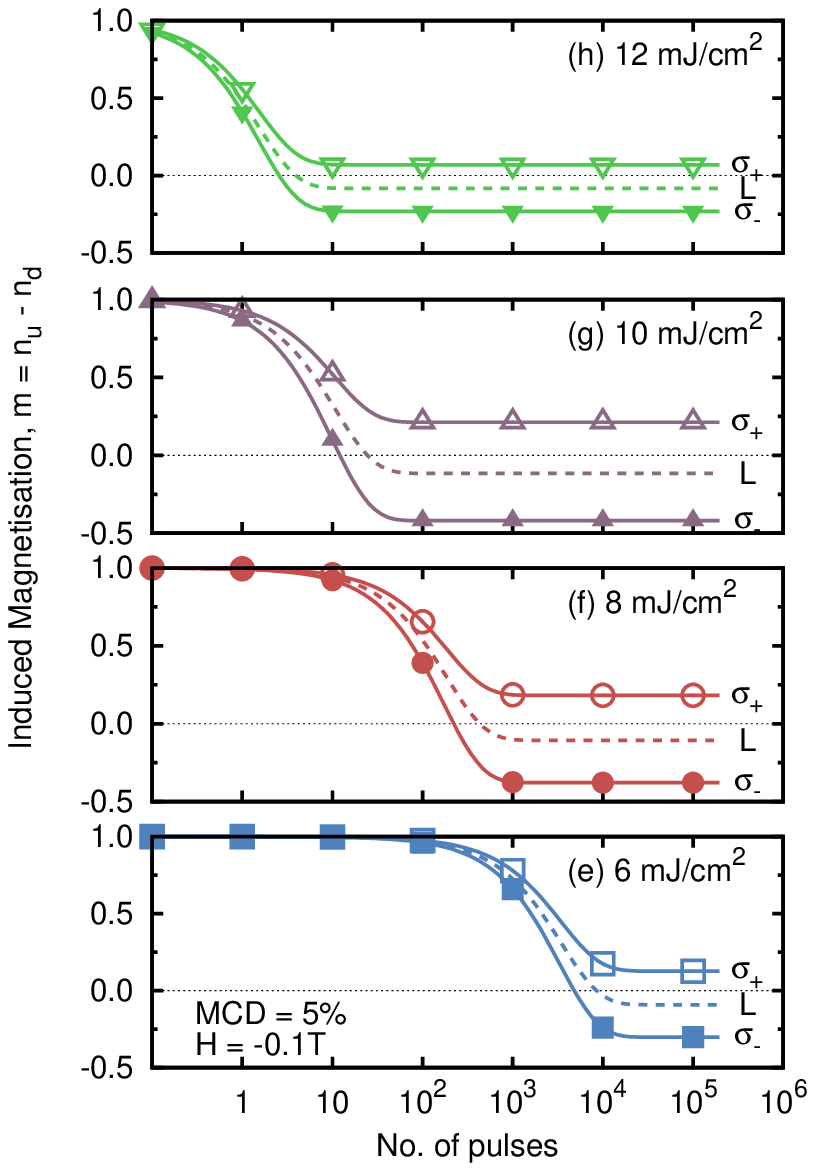}
  	\caption{
    \textbf{Helicity dependent evolution of the magnetization over a series of laser pulses.} The time evolution of the magnetization starting from an initially fully saturated state with a MCD ratio of 5\%. Left column (a-d), zero applied field, right column (e-h), applied field = 0.1T applied in the negative sense relative to the initial magnetization. }
  	\label{fig:2state_mvtime_saturation}
\end{figure}

The effect of an applied field is to bias the thermal switching as shown in figure \ref{fig:induced_mag}.(b) and (c). The experiments by Lambert \etal\ show that a \SI{700}{Oe} field can counter the all-optical switching. Figure \ref{fig:induced_mag} shows the resulting induced magnetization for the case where the applied field is parallel to the magnetization of (b) the low absorption states and (c) the high absorption state. As (b) shows the field and the MCD switching cooperate so the field increases the induced magnetization. In (c) the field counteracts the effect of the MCD switching, reducing the induced magnetization. For this case the field reduces the induced magnetization to approximately 0 for for an MCD close to 2\%. This agrees with the experimental results, showing that a small field can eliminate the effects of the thermal MCD switching.

Finally we consider the effect of the helicity of the laser pulse on the evolution of the magnetization starting from a state of full magnetic saturation. The results are given in Fig.~\ref{fig:2state_mvtime_saturation}, which shows the variation of the magnetization with the number of laser pulses for the two states of helicity. In Fig.~\ref{fig:2state_mvtime_saturation} the left column shows the evolution of the magnetization in zero applied field, and right column, the response to an applied field of 0.1T. For comparison the case of linearly polarized light is also given in each case. In the case of zero applied field the magnetization evolves to zero for the case of linear polarization as expected. For circularly polarized light the magnetization initially decreases with an asymptotic approach to equilibrium values whose sign is dependent on the helicity of the polarized light. The time to equilibrium decreases with increasing fluence (a-d) while the equilibrium value first increases and then decreases. The application of a magnetic field breaks the symmetry as shown in Fig.~\ref{fig:2state_mvtime_saturation}.(e-f). The field, in the sense applied in Fig.~\ref{fig:2state_mvtime_saturation}, assists magnetization reversal, shifting the magnetization more negative for all polarizations of the laser. The results are consistent with the experimental results of Takahashi \etal\cite{Takahashi2016}

\section*{Discussion}

In this study, the underlying physics of optically induced switching in L$1_0$ FePt media has been investigated considering two alternative mechanisms. Firstly, switching triggered by a combination of elevated temperatures and an assumed opto-magnetic field induced by the inverse Faraday effect has been investigated. By using atomistic spin dynamics parameterized from \emph{ab initio} calculations the switching window is seen to require fields that are either of magnitude in excess of 60 T or a duration greater than 5 times that of the laser pulse. 
Such large fields are perhaps justifiable for a 2-sub-lattice ferrimagnet such as GdFeCo, where switching is driven by a 2-magnon bound state~\cite{Barker2013} involving fields of the order of the exchange interaction, however, it seems less likely for a ferromagnet such as FePt. 
A simpler explanation of the results presented by Lambert \etal\ is that of an optically-induced reptation. The thermal activation during the demagnetization caused by the laser allows the grains to switch, and  if the different orientations absorb different amounts of energy from the laser due to the MCD effect then there is a difference in the transition rates. In a simple 2 state Master equation approach, using single-shot transition probabilities determined by the atomistic model, these different transition rates are shown to give rise to an induced magnetization over repeated cycling of the laser. An MCD of 3\% is sufficient to induce a magnetization similar to that seen experimentally. The effect of an applied field is to bias the transition probabilities and as such can then inhibit the reptation effect when it is parallel to the magnetization of the high absorption state or switch the magnetization when in the opposite orientation. This model of the helicity dependent all-optical switching in FePt seems physically justifiable and requires only the assumption of an MCD value of $\Delta \approx 2\%$. One can estimate the MCD from measurements of the polar Kerr rotation in radians which, for L1$_0$ FePt with a 800 nm laser, appears to be approximately 0.87\%\cite{Sato2004}. However, it does not exclude the IFE from playing some role in the switching but importantly it removes the necessity to invoke the unreasonably large opto-magnetic fields required for the IFE to be the dominant mechanism. 
Finally, we have shown that optically-induced switching via a series of laser impulses, the optical reptation effect, is a new mechanism distinct from TIMS in granular ferromagnets. Although the effect is relatively weak, as found experimentally by Lambert \etal\cite{Lambert2014a}, it is likely that optimization of material parameters using the ideas presented here could lead to a new approach to energy assisted magnetization reversal. The method would require temperatures below the Curie temperature which would be a considerable practical advantage.

\section*{Methods}

\subsection*{Atomistic Spin Dynamics}
Atomistic spin dynamics models the magnetic material as Heisenberg (classical) spin moments that are localized to the atomic sites. The dynamics modeled by the time integration of the Landau-Lifshitz-Gilbert equation:

\begin{equation}
\frac{\partial \mathbf{S}_i}{\partial t} = - \frac{ \gamma}{(1 + \lambda^2)\mu_s} \mathbf{S}_i
\times
		\left( \mathbf{H}_i + \lambda \mathbf{S}_i \times \mathbf{H}_i \right)
		\label{eqn:llg}
\end{equation}

where the effective field is given by $\mathbf{H}_i = - \partial \mathcal{H} /\partial \mathbf{S}_i + \boldsymbol{\xi}_i $, $\lambda$ is the atomistic damping (coupling) and $\mu_s$ is the magnitude of the spin magnetic moment. The ensemble is kept at 
a finite temperature by using a Langevin thermostat, whereby a stochastic thermal noise term is coupled into the effective field. The thermal noise is a Gaussian white noise process with the following mean and variance:

\begin{align}
	\langle \xi_{i\alpha}(t) \rangle & = 0 \\
	\langle \xi_{i\alpha}(t) \xi_{j\beta}(s) \rangle & = 
\frac{ 2 \mu_s \lambda k_B T}{\gamma} \delta_{ij} \delta_{\alpha \beta}
\delta(t-s) \label{eqn:noise2moment}
\end{align}

To integrate the coupled equations of motion the semi-implicit method is used\cite{Mentink2010, Ellis2012} which is integrated using a time step of $\Delta t = \SI{1e-16}{s}$. To accelerate the dynamics the model is implemented on graphics processing units (GPUs). 

FePt is modeled in the ordered L$1_0$ phase using an effective,
classical spin Hamiltonian which was constructed earlier
\cite{Mryasov2005} on the basis of first-principles calculations of
non-collinear configurations calculated using constrained local spin
density functional theory.
It was found that
magnetic interaction parameters are strongly affected by the fact that
the magnetic moment of the Pt sites is entirely due to the exchange
fields provided by the Fe sites. It was shown that this important
feature of the electronic interactions can be taken into account
within a model of localized Fe magnetic moments with modified
effective magnetic interactions. In addition to the commonly
considered Heisenberg exchange and single-ion anisotropy these
modified effective magnetic interactions include an isotropic exchange
term dependent on the Pt intra-atomic exchange as well as an effective
exchange mediated two-ion anisotropy \cite{Mryasov2005}.  
The full Hamiltonian, described in detail in Ref.~\onlinecite{Mryasov2005},
is
\begin{equation}
  {\cal H} = - \sum\limits_{i<j} \big( J_{ij} {\bf S}_i \cdot {\bf
               S}_j + d^{(2)}_{ij} S^z_i S^z_j \big) - \sum\limits_i
               d^{(0)} (S^z_i)^2  - \sum_i \mu_s {\bf S}_i \cdot \mathbf{H}
     \label{e:hamiltonian}
\end{equation}
where the ${\bf S}_i = {\bf \mu}_i/\mu_{\mathrm s}$ are
three-dimensional reduced magnetic moments of unit length. The first
sum represents the exchange energy of magnetic moments and the
two-ion anisotropy is not restricted to nearest-neighbor
interactions.  The exchange interactions $J_{ij}$ (and consequently
also $d^{(2)}_{ij}$) have to be taken into account up to a distance
of 5 unit cells until they are finally small enough to be neglected.
The two-ion anisotropy parameters $d^{(2)}_{ij}$ are the dominant
contribution to the uniaxial anisotropy energy in relation to the
single-ion term $d^{(0)}$ which is represented in the second sum.

\subsection*{Incorporating the femto-second laser}

The effect of the laser is incorporated using the two temperature model\cite{Anisimov1974} where the electrons and phonons exist as distinct heat baths in quasi-equilibrium. The laser couples directly to the electron heat bath which then transfers energy to the phonon and spin system. In this case the spin thermal noise above is coupled to the electron temperature and the spin ensemble itself represents the spin heat bath and no separate temperature is assigned to it. The temporal evolution of the
electron, $T_e$, and lattice temperatures, $T_l$, are
governed by the following equations\cite{Mendil2014}:
\begin{align}
	C_e \frac{ \partial T_e}{ \partial t} &=  - G_{el} ( T_e - T_l ) + P(t) - C_e
\frac{ T_e - T_{\text{room}}}{\tau_{ph}} \label{eqn:TE}\\
C_l \frac{ \partial T_l}{ \partial t} & =   G_{el} ( T_e - T_l ) \label{eqn:TL}
\\
P(t) & =  I_0 F \exp\left( -\left( \frac{t-t_0}{\tau_l} \right)^2 \right),
\end{align}
where $C_e$ and $C_l$ are the electron and phonon heat capacities respectively and $G_{el}$ the electron-phonon coupling constant. $P(t)$ describes the laser heating power which has a Gaussian shape centered at $t_0$ with the pulse width $\tau_l = 100 \si{fs}$. The fluence, $F$, is coupled through a material specific constant $I_0$. Parameters for granular FePt are given by Mendil \etal\cite{Mendil2014} which are extracted from comparison to experimental results.

Theoretical treatments of the inverse Faraday effect predict that in interaction of the polarized light with a magnetized media will create a magnetization or magnetic field within the media\cite{VanderZiel1965, Hertel2006}. Following the description given by Kimel \etal\cite{Kimel2007} the resulting field depends on the electric field of the laser, $\mathbf{E}$, as

\begin{equation}
	\mathbf{H}_\text{IFE} = - \varepsilon_0 \alpha \left[ \mathbf{E} \times \mathbf{E}^*\right],
\end{equation}

where $\varepsilon_0$ is the vacuum dielectric constant and $\alpha$ is the coefficient of the magnetization linear term in the expansion of the dielectric tensor $\varepsilon_{ij}$. The direction of the field is either parallel or anti-parallel to the propagation direction, depending on the helicity, and so we consider perpendicular set up where this is along the film normal, i.e. $\hat{\mathbf{z}}$. Whilst there is various theoretical predictions for the strength of this field we leave this as an open parameter. Thus, the opto-magnetic field utilized in the spin dynamics simulations is 

\begin{equation}
  \mathbf{H}_\text{IFE} = h_{\text{IFE}} \, \hat{\mathbf{z}} \begin{dcases}
			  \exp( -((t-t_0)/\tau_l)^2 )  & t < t_0\\
			 1 & t_0 < t < t_0 + t_d\\
              \exp( -((t - (t_0 + t_d) )/\tau_l)^2 )  & t > t_0 + t_d \\
			 \end{dcases}
\end{equation}

where $h_\text{IFE}$ and $t_d$ are the magnitude and duration of the IFE field respectively. $t_0$ and $\tau_l$ are the centre and pulse width of the laser in the same manner as used to model the heating effects of the laser.

\subsection*{Probability Of Switching Extracted From Atomistic Spin Model}

To parameterize the two state Master equation the probability of a single grain switching is calculated using atomistic spin dynamics. The results are shown in figure \ref{fig:Switching_H0}; and to provide a functional form the following equation is fitted to each of the data sets

\begin{equation}
P(F) = \frac{P_\infty}{2}\left( 1 + \tanh \left( \frac{(F-F_0)}{\Delta F} \right) \right), \label{eqn:P_switch}
\end{equation}

with $P_\infty$, $F_0$ and $\Delta F$ as fitting parameters. With no external field the effect of a sufficiently high fluence will be to demagnetize the grain completely and as $F \rightarrow \infty$ the grain will have an equal chance of entering either orientation meaning $P_\infty = 1/2$. With a constant external field the switching will be biased but thermal effects mean that not all of the grains will align with the field as the system cools implying that $P_\infty$ for $H \neq 0$ will only tend to 1 for high fields. Since fitting was done for zero field and $\pm$ 1 T to model the effects of a smaller field a linear interpolation of the constants in equation \eqref{eqn:P_switch} is used.


%

\section*{Acknowledgements}
This work made use of the facilities of N8 HPC provided and funded by the N8 consortium and EPSRC (Grant No. EP/K000225/1) co-ordinated by the Universities of Leeds and Manchester and the EPSRC Small items of research equipment at the University of York ENERGY (Grant No. EP/K031589/1). Work at UCSD was supported by the ONR MURI program. We would like to thank R. F. L. Evans, T. A. Ostler, Y.K Takahashi and V. Lomakin for fruitful discussions.

\section*{Author contributions statement}

E.F. and R.C. conceived the initial investigation. M.E. performed the atomistic spin dynamics simulations. M.E. and R.C. devised the two state Master equation model of the induced magnetization and prepared the manuscript. All the authors analyzed the results and reviewed the manuscript.

\section*{Additional information}
\textbf{Competing financial interests}: The authors declare no competing financial interests.

\end{document}